# Overview of Caching Mechanisms to Improve Hadoop Performance

Rana Ghazali [a,b], Douglas G.Down [b]
Ghazalir@mcmaster.ca, Downd@mcmaster.ca

a: Department of Computer Engineering, North Tehran Branch, Islamic Azad University, Tehran, Iran
b: Department of Computing and Software, McMaster University, 1280 Main St W, Hamilton, ON, Canada

## Abstract

In today's distributed computing environments, large amounts of data are generated from different resources with a high velocity, rendering the data difficult to capture, manage, and process within existing relational databases. Hadoop is a tool to store and process large datasets in a parallel manner across a cluster of machines in a distributed environment. Hadoop brings many benefits like flexibility, scalability, and high fault tolerance; however, it faces some challenges in terms of data access time, I/O operation, and duplicate computations resulting in extra overhead, resource wastage, and poor performance. Many researchers have utilized caching mechanisms to tackle these challenges. For example, they have presented approaches to improve data access time, enhance data locality rate, remove repetitive calculations, reduce the number of I/O operations, decrease the job execution time, and increase resource efficiency. In the current study, we provide a comprehensive overview of caching strategies to improve Hadoop performance. Additionally, a novel classification is introduced based on cache utilization. Using this classification, we analyze the impact on Hadoop performance and discuss the advantages and disadvantages of each group. Finally, a novel hybrid approach called Hybrid Intelligent Cache (HIC) that combines the benefits of two methods from different groups, H-SVM-LRU and CLQLMRS, is presented. Experimental results show that our hybrid method achieves an average improvement of 31.2% in job execution time.

***Keywords: Hadoop performance, Caching mechanism, MapReduce, Hybrid Intelligent cache***

## 1-Introduction

Hadoop [1] is an open-source framework for the parallel processing of large datasets in a distributed environment. Hadoop consists of two major components: HDFS (Hadoop Distributed File System) [2] as a storage media and MapReduce [3] as a parallel processing model. HDFS has a master/slave architecture, dividing input data into data blocks with identical sizes that are distributed among data nodes. Moreover, it provides multiple copies of each data block (referred to as the replication factor) and keeps them in different data nodes to provide high fault tolerance. MapReduce is a parallel programming model that includes two types of tasks: Map and Reduce. In the first phase, the Map task converts input data into <key, value> pairs (referred to as intermediate data). These intermediate data are then sorted and shuffled based on identical keys. In the second phase, these data are passed to the Reduce task as input data and values with the same keys are merged to generate final results.

Hadoop faces some challenges that have a significant impact on reducing its efficiency:

1. One of the bottlenecks of Hadoop is related to HDFS because it is based on a hard disk drive (HDD) system, which has high access time for I/O operations, resulting in a negative impact on the overall job execution time.
2. The shuffling phase in MapReduce is a time-consuming operation that can account for more than 33% of the job execution time.
3. A large amount of intermediate data is discarded after finishing the process. Due to the lack of any mechanism to identify duplicate computations in MapReduce, it is necessary to reload and recompute intermediate data whenever applications need to reuse them, leading to resource wastage and poor performance.
4. In an iterative program, data may be unchanged from one iteration to the next. MapReduce does not have any mechanism to reuse computational data so data must be re-loaded and re-processed at each iteration. In addition, recognizing termination conditions or fixed points (when the application's output does not change for consecutive iterations) requires an extra MapReduce job on each iteration. These two issues lead to reducing the effective use of resources such as network bandwidth and CPUs.
5. The job scheduler may assign tasks to a node that does not contain the required input data. In this case, data must be transferred from a remote data node, resulting in increased network traffic and delays in execution.
6. When failure occurs, only the requested data block is recovered and is not saved for future use. Therefore, if this data block is needed again, the recovery operation must be repeated.

In recent years, many researchers have proposed caching mechanisms as an approach to tackle these challenges. The caching mechanism could prefetch on-demand data into cache memory, which has a more rapid access time than HDFS. As a result, an effective caching strategy can have a significant effect on reducing execution time and improving Hadoop efficiency. A caching mechanism consists of two phases: the placement phase and the delivery phase. The placement phase determines how to store and place data into the cache memory, based on their popularity. The second phase is the delivery phase, which retrieves data from cache memory based on the actual demands of the tasks. In this phase, the main limitations are the rate to serve the requested content and network congestion. A caching mechanism must cope with restricted cache space. For instance, a replacement algorithm is needed to determine which existing data should be evicted when inserting new content if the cache is at full capacity.

In this paper, we provide a comprehensive overview of proposed methods that utilize a caching mechanism for improving Hadoop performance. We classify them based on the performance issue that they address, and we analyze their effect on Hadoop performance and discuss their advantages and disadvantages. Also, we present a novel hybrid approach called Hybrid Intelligent Cache (HIC) that combines the benefits of two methods from different classes, H-SVM-LRU and CLQLMRS.

The rest of this article is organized as follows. Section 2 discusses the impact of using a caching mechanism on Hadoop performance. We classify existing caching solutions based on their approach to solving Hadoop's performance issues in Section 3. We compare them and investigate their effect on Hadoop performance in Section 4. The novel hybrid method which combines H-SVM-LRU with CLQLMRS is presented in Section 5 and finally, Section 6 contains concluding remarks and suggestions for future work.

## 2- The benefits of a caching mechanism to address Hadoop performance challenges

The Hadoop framework has been found to suffer from a number of problems that result in poor performance. To address these issues, many researchers have suggested the use of caching mechanisms. This strategy offers several advantages when applied to Hadoop, including:

1. By caching data, access time is significantly reduced compared to accessing data from a disk. This leads to a reduction in the number of I/O operations, a decrease in prefetching delays, and an overall improvement in execution time. In this way, caching can help to solve the problem of I/O operation time.
2. Storing intermediate data in the cache can prevent resource wastage and eliminate the need for recomputing in the MapReduce programming model, resulting in decreased execution time and resource utilization.
3. By reducing the amount of data transmission through the network, caching data has a positive impact on reducing execution time.
4. When assigning tasks to nodes, considering cache locality in scheduling decisions can improve data locality, resulting in improved system performance.

Given the benefits of using caching mechanisms to improve Hadoop performance, there has been considerable research in this area. We present a novel classification based on the specific Hadoop problems that existing proposed caching mechanisms have been designed to solve. One benefit of this classification is that it identifies where there is the potential to combine caching mechanisms from different groups. We give a concrete example of this in Section 5.

## 3- Novel classification of caching mechanisms in Hadoop

As mentioned in the previous section, Hadoop can encounter challenges for which a caching mechanism may provide a solution. Caching mechanisms have been the subject of a significant body of research. These can be divided into two primary categories contingent upon the principal research objectives. Within the initial category, the focus is on the strategic deployment of cache resources and the identification of data types amenable to caching, thereby addressing foundational Hadoop performance issues. This particular classification of caching mechanisms shall be referred to as the "basic caching mechanism" within the context of this article. The second category encompasses efforts geared towards the enhancement of various facets inherent to caching mechanisms. For example, the refinement of cache replacement policies to optimize the utilization of limited cache capacity while circumventing issues of cache contamination may yield improvements for cache hit ratios. Furthermore, improving prefetch mechanisms by orchestrating the prefetching of a suitable volume of data at opportune junctures has the potential to positively influence data retrieval time. Consequently, the nomenclature "enhanced caching mechanism" will be used for this latter grouping. In this study, a novel classification of basic caching strategies is presented based on the performance issue that they are designed to address. These caching mechanisms are categorized into four groups, namely:

1. Caching mechanisms for improving data access time
2. Caching mechanisms for avoiding duplicate computations
3. Caching mechanisms for decreasing data transmission time
4. Caching mechanisms for other purposes.

Furthermore, enhanced caching mechanisms can be divided into two groups based on their approach to improving performance:

1. Enhanced caching via improving the cache replacement strategy
2. Enhanced caching via optimizing the prefetch mechanism.

This paper will discuss existing caching strategies in each group, along with their strengths and weaknesses.

### 3.1 Basic caching mechanisms

This section focuses on the implementation of caching mechanisms to optimize the performance of Hadoop. As previously mentioned, leveraging caching strategies in Hadoop presents several advantages that can be categorized into four distinct groups. In the subsequent sections, we will delve into a range of caching approaches, within their respective groups.

#### 3.1.1 Caching mechanisms for improving data access time

In Hadoop, data is stored in the Hadoop Distributed File System (HDFS), which is specifically designed to handle large datasets across multiple nodes in a cluster. However, since HDFS is based on a disk mechanism, accessing data from a disk can be a time-consuming process. As a result, caching frequently accessed data in memory can speed up data access by reducing the number of referrals to HDFS to read data from disk. This can significantly reduce data access time and improve overall system performance. In this section, we explore some caching strategies that are commonly used to achieve this purpose.

One approach proposed by Zhang et al. is the *HDFS-based Distributed Cache System (HDCache)* [4], which is a data access model designed for write-once-read-many files. This model comprises a client library (*libCache*) and multiple cache services. The *libCache* library contains shared memory access and the communication and control module, which interacts with the *HDCache* service on the same host, communicates with *ZooKeeper* servers remotely, and calculates hash values of desired files to locate a specific cached file. The multiple cache services provide three access layers, including an in-memory cache, a snapshot of the local disk, and the actual disk view. This method can improve real-time file access performance in a cloud computing environment. However, it does not consider complex and dynamic data access models of real-time services.

Another caching middleware is *Hycache* [5], which serves as an intermediary between the distributed file system and local storage devices to enhance SSD performance at a low cost. *Hycache* consists of three components: the Request Handler, File Dispatcher, and Data Manipulator. The Request Handler receives requests from HDFS and passes them to the File Dispatcher, which uses cache replacement algorithms like LRU and LFU to decide which files in the SSD should be swapped to the HDD. Data are manipulated between two access points via a Data Manipulator. The merits of this method are low latency in data access, high throughput, and multithread support. However, it is not scalable and only considers local data movement.

Hycache+ [6] presents a scalable caching middleware on the computing node side to improve I/O performance for the parallel file system. Additionally, the 2-Layer Scheduling (2LS) approach was proposed to enhance cache locality. In the first layer, the job scheduler assigns a job to available nodes to minimize file size transmission through nodes. In order to maximize the total cached file size, the second layer locates data across local storage by considering file size and its access frequency. Hycache+ uses a heuristic caching approach instead of LRU to optimize caching effects. Minimizing network cost and scalability are advantages of this strategy.

The *Two Level greedy caching strategy* [7] is a means to address the discrepancy between disk access time and bandwidth in large cluster-based systems. The primary objective is to develop a proactive fetching and caching mechanism by integrating the Memcached distributed caching system with Hadoop. The Two-Level greedy caching approach combines two distinct simple greedy caching policies, namely receiver-only greedy caching and sender-only greedy caching. This method aims to enhance data access time by augmenting the cache hit ratio. However, it is important to note that this technique also introduces additional overhead in terms of network traffic.

### 3.1.2 Caching mechanisms for avoiding duplicate computations

In certain MapReduce applications, particularly those that involve iterative processes, it may be necessary to reuse intermediate data generated during the Map phase. Failure to save these data may result in redundant computation, which can be a significant waste of resources. By caching such data, duplicate calculations can be eliminated and the data can be easily retrieved from the cache, thereby saving time. Consequently, caching plays a crucial role in Hadoop as a means of avoiding the recomputation of intermediate data. Storing such data in a cache enables quick and easy access, thereby improving performance and reducing processing time. In the following, we discuss various caching strategies that can be employed to address this issue.

*Haloop* [8] was specifically developed to address the challenges posed by iterative applications in the Hadoop framework. To achieve this goal, *Haloop* employs caching of invariant data that can be reused in subsequent iterations, with the cache being indexed to expedite processing. Three types of caches are utilized in this approach. First, the Mapper input cache is used to avoid non-local data reads during non-initial iterations in the Map phase. Second, the Reducer input cache is employed to reduce shuffling time. Third, the Reducer output cache is utilized to identify a fixed point where the application's output is unchanged for two consecutive iterations. In addition, the Loop Aware Task scheduler was proposed to consider inter-iteration locality and allocate Map and Reduce tasks to worker nodes that access the same data but in different iterations. To achieve this, the NameNode must provide a mapping between the data block and the worker node that processed them in the previous iteration. While this strategy can significantly reduce run time and shuffling time, it requires a fixed number of Reduce tasks across iterations.

*Incoop* [9] was designed specifically for incremental applications to identify changes in input data and automatically update the corresponding output by reusing previous results. This approach employs three techniques to achieve its objectives. First, the Incremental HDFS technique provides stable partitions of input data by using content-based chunking instead of fixed-size chunking to maximize the overlap between data chunks. Second, the Contraction phase controls the granularity of tasks in terms of their size and dependencies by combining Reduce tasks hierarchically. Third, a memorization-aware scheduler was developed to minimize data transmission across networks by assigning tasks to machines that store the results of sequential jobs for reuse. This policy considers the tradeoff between maximizing the locality of results and minimizing straggler tasks by employing a work-stealing algorithm that ensures nodes are not idle while runnable tasks are waiting. *Incoop* is suitable for compute-intensive jobs as it avoids recomputation by reusing results, and it can reduce execution time through location-aware scheduling for memorization.

However, this approach can generate computational overhead in the incremental HDFS for small dataset sizes.

*Data-Aware Cache (Dache)* [10] was proposed as a mechanism to avoid discarding a large amount of intermediate data and recomputing them, thereby improving CPU utilization and accelerating execution time. To achieve this goal, results of tasks are cached, and each task queries the cache before its execution. Additionally, Dache includes two types of cached items stored in a Map cache and a Reduce cache. This method employs a novel cache description scheme that represents each cached item by a tuple comprising the input file and an operation applied to it. Furthermore, a request and reply protocol is utilized to access a cached item by matching the input file and its operation to reuse them. While this strategy provides benefits such as reducing job completion time and improving CPU utilization, there is a limitation in the data partition scheme such that it should use the same data split in both the data and cached item. This technique is well-suited for incremental processing that requires the application of duplicate calculations.

*Redoop* [11] was designed to optimize recurrent queries in the Hadoop infrastructure through the use of window-aware techniques. This includes adaptive window-aware data partitioning, cache-aware task scheduling, and an inter-window caching mechanism. The dynamic data packer splits input data partitions into smaller panes based on statistical information gathered by the execution profiler, while the inter-window caching mechanism avoids redundant computation by caching intermediate data. Additionally, cache-aware task scheduling performs load balancing by utilizing cache locality in its decisions. *Redoop's* strengths include reducing I/O costs and providing workload balancing, but its weakness lies in the generation of overhead due to the gathering of statistical information.

*CURT MapReduce* was proposed in [12] with the ability to cache and utilize task results to avoid recomputation overhead. The strategy involves the use of the *Intelligent Square Divisor (IDS)* algorithm to split input and intermediate data into a specific number of pieces, a Cache Data Seeker to search for existing input data and corresponding results from previous tasks, and a Cache Data Creator to format cached data into input-to-output tuples. *CURT MapReduce* is appropriate for iterative applications as it reduces resource wastage and computation overhead, leading to a positive impact on performance. However, it may not be efficient for applications with a low volume of recomputation.

#### 3.1.3 Caching mechanisms for decreasing data transmission

The utilization of caching mechanisms is a viable approach to improving the efficiency of data transmission across networks. This is achieved by taking into account the cache locality in job scheduling policies. Consequently, tasks can be allocated to nodes that have cached the requisite data, leading to a reduction in data access time, and an increase in task locality rates. By doing so, this approach effectively prevents the need for extra data transmission through the network, contributing to overall improved Hadoop performance.

Zhang et al. [13] employed a weighted bipartite graph paradigm in which vertices correspond to Map tasks and resources, and the relationships between these elements are established through edge weights. Within this framework, Map tasks undergo prioritization and are subsequently organized into a selected matrix predicated on the spatial arrangement of task input data. The selection of suitable worker nodes for task processing is facilitated through the implementation of maximum matching algorithms. While this approach improves data and cache locality for map tasks, it does not incorporate considerations pertaining to the overall workload of the cluster.

*LARD (Locality Aware Request Distribution)* [14] leverages disk buffer caches and predicts the storage location of cached data by using information about where previous requests were processed. The effectiveness of this approach decreases for large file sizes and frequent alterations, where more prediction errors result in increased disk access durations and suboptimal system performance.

Embedded within the operating system's buffer cache, *CATS (Cache Aware Task Scheduling)* [15] is characterized by two primary components: a buffer cache probe that gathers information regarding cached data across each worker node, and a task scheduler that considers cache locality when making scheduling decisions. A drawback of this approach is that data locality is not considered. This drawback is further exacerbated by the overhead resulting from disk accesses that ensue when cache-local tasks cannot be instantiated.

The *Adaptive Cache Local Scheduling Algorithm (ACL)* [16] is based on a cache affinity-aware replacement algorithm intended for data block eviction from the in-memory cache. To determine cache affinity, a value, denoted as C, is computed, giving the number of times a task is overlooked by the job scheduler in order to satisfy cache locality. The value of C is proportional to the percentage of cached input data for the job. To mitigate instances of starvation, the algorithm dictates that if the scheduler overlooks a task D times, said task should be dispatched to a node containing the requisite input data and possessing an available slot. Nevertheless, this strategy does not factor in the performance ramifications resulting from concurrently executing applications within a given workload. Moreover, the *ACL* algorithm has overhead for scheduling and deployment on cache-local and data-local nodes. Furthermore, the efficacy of this approach requires the tuning of parameters C and D.

In an effort to enhance task locality in terms of both cache locality and data locality, Ghazali et al. proposed *CLQLMRS (Cache Locality with Q-Learning in MapReduce Scheduling)* [17]. This scheduling method employs a form of reinforcement learning known as Q-Learning to train a scheduling policy without requiring prior environmental information. The objective of this approach is to improve execution time by reducing the amount of data transmission. *CLQLMRS* is particularly suitable for I/O-Bound jobs and data-oriented applications.

### 3.1.4 Caching mechanisms for other purposes

Despite its original intended purpose, caching mechanisms in Hadoop have been utilized by researchers for other beneficial purposes, such as enhancing scalability, reducing shuffling time, and improving data recovery. In the following section, we describe these approaches in greater detail.

The *Separation* method [18] was developed to address the scalability issues of Hadoop clusters and overcome the NameNode's memory limitations when storing filesystem metadata. To achieve this, the NameNode utilizes a cache to retain frequently used data for high availability. Furthermore, a maximum memory capacity threshold is set for the storage of metadata in the NameNode. Once the volume of metadata in the NameNode's memory reaches a threshold value, the separation algorithm is activated. This algorithm transfers the least recently used metadata from the NameNode to secondary storage, thereby caching only frequently used metadata. The advantages of this strategy are improved availability and scalability, although it requires adaptation of the threshold value, and generates some overhead by introducing count and last-time fields in the metadata.

Maddah Ali et al. proposed *Coded MapReduce* [19] to reduce the communication load during the shuffling phase. This approach employs coded multicast to exploit the repetitive mappings of the same data block at different servers. Although this method increases processing time due to the repetitive execution of Map tasks, it effectively balances the computation load with the inter-server communication load. Coded MapReduce can reduce communication load by a multiplicative factor that grows linearly with the number of servers. However, this approach imposes additional calculations, which can negatively impact execution time.

When a failure occurs in a Hadoop cluster, and some DataNodes become unavailable, the degraded process begins to serve incoming data requests by utilizing their replicas in surviving nodes. However, requested data can only be recovered and not shared, leading to wastage of resources, particularly network bandwidth, and resulting in increased execution time and poor performance. To address this issue, *Cooperative, Aggressive Recovery and Caching (CoARC)* [20] was introduced as a novel data recovery mechanism in the distributed file system. This approach aims to avoid redundant recovery of failed blocks by recovering unavailable data blocks on the same strip in addition to the requested data and then caching them in a separate node for accessibility. In this mechanism, the Least Recently Failed (LRF) algorithm was presented as a new cache replacement algorithm. As most of the DataNodes come back alive after some time, there is no need to write back the data blocks to HDFS. The benefit of this approach is that it recovers all unavailable data blocks in the strip without any additional overhead and network traffic, leading to reduced execution time.

### 3.2 Enhanced caching mechanisms

A considerable amount of research has been conducted to enhance caching mechanisms, either by improving cache replacement or prefetch strategies. Consequently, there are two groups of enhanced caching mechanisms. This section will provide a detailed survey of both groups.

#### 3.2.1 Enhanced caching via improving cache replacement strategy

The cache replacement algorithm plays a vital role in improving the cache hit ratio and cache space utilization. Various cache replacement techniques consider different features of the cached items to evict them when the cache capacity is reached. This section will investigate different cache replacement strategies employed in Hadoop, including their advantages and disadvantages.

Within the context of *PacMan* [21], the concurrent execution of parallel jobs is organized in a wave-like fashion, characterized by the all-or-nothing property. This operational framework incorporates two innovative in-memory cache replacement strategies, namely, *LIFE* and *LFU-F*. The *LIFE* strategy is distinguished by its eviction of data blocks associated with files possessing the widest wave-width, a tactic designed to diminish the average job completion time. Conversely, *LFU-F* evicts data blocks with infrequent access, thereby optimizing cluster efficiency. Notably, both mechanisms circumvent cache pollution through the implementation of a window-based aging mechanism and prioritize caching data blocks originating from completed files. These strategies are most suitable for data-intensive parallel jobs.

The *WSClock* algorithm, as introduced in *Enhanced Data-Aware Cache (EDACHE)* [22], is a cache replacement algorithm that relies on a circular list for the management of cached items, particularly intermediate data. The determination of evicted data is based on the examination of a reference bit and the last time of usage. Specifically, the reference bit is initially inspected; if its value is found to be one, indicative of recent usage, the reference bit is reset, the timestamp of last usage is updated, and the clock hand progresses. Conversely, in cases where an item's age surpasses a predefined threshold value, it is selected for eviction. This algorithm may not be well-suited for managing large blocks of data due to the extended search times required to locate requested content.

*A Modified ARC replacement algorithm* was introduced in [23]. This algorithm integrates the Least Recently Used (LRU) and Least Frequently Used (LFU) approaches, resulting in the creation of two distinct cache segments: the recent cache and the frequent cache. These segments store data blocks, each equipped with an individual history section containing references to the corresponding data blocks. These references play a pivotal role in determining eviction decisions. When a request for a specific block is made, the algorithm examines the references within both history caches. If the references are found, the corresponding block is allocated within either the recent cache or the frequent cache. Conversely, if the references are absent, the cache consults the history caches and subsequently accommodates the request from one of them. This process enhances caching efficiency, expediting the retrieval of files for preliminary assessments. A reference discovery in the recent history prompts the placement of the block in the recent cache, followed by its transfer to the frequent cache. As such, successful references within either history cache trigger the removal of the reference itself and facilitate the placement of the corresponding block within either the recent or frequent cache. Importantly, the caching of a block encompasses the concurrent caching of associated metadata. When either of the caches reaches full capacity, the eviction of a block is executed from the recent or frequent cache; however, the pertinent reference is retained within the corresponding history. When saturation occurs within either of the history caches, the references simply exit the cache, thereby concluding their presence.

Another noteworthy cache replacement algorithm is the *Cache Affinity Aware Cache Replacement Algorithm* [24]. This novel approach leverages cache affinity as a quantifiable metric for prioritizing the caching of input data. In cases where multiple data blocks exhibit identical cache affinity, the algorithm employs the LRU policy to determine the eviction of a particular block. The efficacy of this method hinges upon the availability of accurate information pertaining to the cache affinity of various applications.

The block goodness-aware cache replacement strategy [25] employs a dual-metric framework to effectively administer cache space. Specifically, this approach incorporates the Cache Affinity (CA) metric, representing an application-oriented attribute that measures the degree of affinity exhibited by applications toward cache resources. Concurrently, the Block Goodness (BG) metric measures the intrinsic popularity of distinct data blocks within the cache context. This strategy initially computes the BG value, a process informed by the cumulative access frequency of data blocks. The eviction selection process ensues, wherein a data block possessing the smallest BG value is chosen for removal from the cache. In scenarios where multiple data blocks have the minimal BG value, the data block endowed with the earliest access timestamp is chosen.

The *Adaptive cache algorithm* [26] incorporates a dual cache replacement framework encompassing *Selective LRU-K (SLRU-K)* and *Exponential-Decay (EXD)* mechanisms within the Hadoop Distributed File System (HDFS) cache, specifically designed for Big SQL workloads. Initially, the algorithm partitions tables, directing them into the HDFS cache. Within this context, the *SLRU-K* strategy adopts a weight heuristic to facilitate the selective inclusion of partitions within the cache. This approach requires the maintenance of a list capturing the last access times of the K most recent interactions for each partition. This algorithm necessitates a greater allocation of space to accommodate the storage of access time data, a notable drawback. In contrast, the *EXD* mechanism exclusively retains information concerning the most recent access time, utilizing this data to compute a score for each partition. This score determines the prioritization between access frequency and recency. The Adaptive *SLRU-K* and *EXD* approach dynamically adapt their behavior in response to diverse workload access patterns, adjusting their parameter values to align with the requirements of specific workloads.

Beyond the above approaches, several cache replacement strategies leverage the capabilities of machine learning techniques to improve cache hit ratios and optimize cache space utilization.

The *AutoCache* strategy [27] employs a lightweight gradient-boosted tree model, specifically the *XGBoost* algorithm, to forecast file access patterns within the HDFS cache environment. This predictive model quantifies the likelihood of accessing a particular file through the derivation of a probability score. This probability score guides the cache replacement policy, strategically mitigating cache pollution. Notably, when the available cache space falls to below 10% of its total capacity, the eviction procedure is triggered. This operation persists until the cache's capacity drops below the 85% threshold. A significant advantage of this strategy is its minimal computational overhead, achieved by constraining computations to a predetermined number of files.

To mitigate cache pollution, *H-SVM-LRU* [28] utilizes a smart classification *Support Vector Machine (SVM)* to classify cached items into two classes based on their likelihood of future reuse. Subsequently, the LRU cache replacement strategy is employed, resulting in an improved cache hit ratio. This combined strategy can have a notable effect on job execution time, as more tasks have the opportunity to utilize cached data instead of accessing data from HDFS. Consequently, the reduction in I/O operation time is one of the advantages of this approach. However, it should be noted that the need for training data poses a limitation to the effectiveness of this method.

### 3.2.2 Enhanced caching with prefetching

The implementation of a prefetching mechanism, designed to retrieve data blocks from the HDFS and store them within the cache, can be effective in the reduction of data access latency. To effectively harness the benefits of prefetching techniques, one must account for specific conditions, including the optimal timing for initiating prefetch operations to minimize conflicts with concurrent activities, as well as ensure timely utilization of prefetched data. Moreover, it is essential to prefetch an appropriate volume of data, aligning with the cache capacity, thereby mitigating cache loss ratios. In this section, a number of prefetching mechanisms are described, each aimed at enhancing the performance of the Hadoop framework:

The realm of big data applications is characterized by the manipulation of extensive data block volumes, presenting a challenge in facilitating unhindered access to input data blocks by all computational tasks from the cache. The inherent difficulty lies in the low likelihood of data blocks being accessed from the cache prior to eviction. In response, a solution known as *Just Enough Cache (JeCache)* was introduced [29]. This solution introduces a just-in-time data block prefetching mechanism, which dynamically assesses data block access patterns and computes the average data processing duration to ascertain the minimal number of data blocks that warrant cache retention. *JeCache* is composed of two principal components: (1) Prefetch information generation, which leverages job history logs to identify data blocks deserving of initial caching and establishes a prefetch sequence for data blocks during job execution; and (2) Prefetch controller, responsible for monitoring data block access within each worker node and orchestrating the eviction of data blocks from the cache once their processing concludes. Notably, this approach leads to a reduction in cache resource demands with a resulting positive impact on execution times. However, it is important to highlight that *JeCache* exclusively addresses read-caching scenarios.

Aiming to curtail data transmission latency between remote and processing nodes within a heterogeneous cluster, Vinutha et al. [30] proposed a *proactive prefetch thread*. This thread proactively retrieves requested input data from remote nodes, depositing them in the buffer of the processing node, which serves as a transient storage space. This strategy can reduce job execution times, by overlapping data transmission with processing activities and improving data locality rates during task launches. Nevertheless, it remains noteworthy that despite the implementation of this prefetching strategy, the initial data transmission still necessitates a certain waiting period.

In an alternative approach presented in [31], the *streaming technique* was introduced to facilitate simultaneous data transfer and processing. By harnessing this technique, the reduced data size inherent to streaming alleviates transmission waiting times.

Kalia et al. [32] proposed *speculative prefetching*, a strategy that factors in node processing capacity when loading input data onto processing nodes. This method leverages *K-Nearest Neighbors (KNN)* clustering, utilizing the Euclidean distance metric to group intermediate data. The goal is to bolster data locality rates for Reduce tasks, consequently amplifying overall performance. It is important to underscore, however, that this approach does not account for additional factors such as workload capabilities and DataNode throughput.

In [33], a *two-tiered correlation-based file prefetching mechanism* and dynamic replica selection strategies were introduced. These strategies collectively aim to minimize data access latency and alleviate the burden on overloaded DataNodes through load balancing measures. The approach encompasses four distinct placement patterns tailored to the storage of fetched data.

*Smart prefetch* [34] is an intelligent prefetching mechanism that comprises three steps. In the initial phase, the appropriate prefetch time is determined based on the progress rate of tasks, such as Map tasks and Reduce tasks, taking into account the processing capacity of the nodes. Subsequently, in the second phase, the volume of prefetched data is determined using the *K Nearest Neighbors (KNN)* clustering algorithm, which relies on the Euclidean distance between data blocks. Finally, a data locality metric is employed to determine the placement of the prefetched data. Experimental results indicate that this mechanism has a significant impact on performance by enhancing data locality and facilitating increased access to data from the cache.

## 4. Comparison of Caching strategies

In this section, we compare the introduced Hadoop caching strategies based on different aspects, including the techniques applied, cache specifications, and their impact on the performance of the Hadoop framework.

### 4.1 Comparison of Caching Characteristics

In this subsection, we address several key questions to analyze the characteristics of caching mechanisms utilized in Hadoop.

1. Which Hadoop cache level is used? We explore the specific cache levels employed in Hadoop, such as Distributed Cache, In-Memory Cache, HDFS Cache, and Memcache. Distributed Cache facilitates the distribution of small, read-only files, archives, or other resource types to nodes within a Hadoop cluster. Its primary purpose is to make these resources available to tasks running on the nodes. The Distributed Cache copies the required files to the local disk of each worker node before executing the corresponding task. The In-Memory Cache, on the other hand, involves storing frequently accessed or computed data in the main memory (RAM), enabling faster data access and processing.

The HDFS Cache is designed to enhance the performance of data access in HDFS. It allows caching of frequently accessed files or portions of files in the memory of DataNodes, which are individual nodes responsible for data storage and processing within a Hadoop cluster. Caching data in the HDFS Cache enables subsequent read requests for that data to be directly served from the cache, reducing disk I/O and improving overall performance. Memcache, an open-source distributed in-memory caching system, is not specific to Hadoop but can also be employed as a caching layer in Hadoop deployments. It stores data in the memory of multiple servers, enabling fast data access and retrieval. Memcache can be utilized in Hadoop applications to cache frequently accessed data or intermediate results, thus enhancing processing speed and efficiency.

2. What are the cached Items? Different types of items can be cached in Hadoop, including files or data blocks. These items can be further categorized based on their data types, such as input data, intermediate data generated during the execution of MapReduce jobs, and the final results of those jobs.
3. Where are the cached items placed? Cached items can be located through client-side caching or server-side caching. Client-side caching involves caching items on the local disk of the client application, which is suitable for smaller data sets. On the other hand, server-side caching involves caching items on the DataNodes within a Hadoop cluster, which is beneficial for larger data sets that cannot fit into the memory of client machines.
4. Which access pattern is used in the cache? The access pattern in the cache can be classified as read-through caching, write-through caching, or write-behind caching. Read-through caching involves retrieving data from the cache if it is present; otherwise, the cache fetches the data from the data source. Write-through caching involves first writing data to the cache and subsequently updating the data source. In contrast, write-behind caching writes data to the cache and asynchronously updates the data source. This mechanism is particularly useful when write operations occur frequently.
5. Which cache replacement algorithm is applied? When the cache reaches capacity, a cache replacement algorithm is employed to evict a cached item and free up space. Commonly used algorithms include Least Recently Used (LRU), Least Frequently Used (LFU), and First In First Out (FIFO) caching, each with its eviction strategy.
6. How to manage the cache? Cache management is a critical aspect when it comes to optimizing performance in Hadoop environments. In this regard, three prevalent strategies are commonly employed: shared cache management, distributed cache management, and centralized cache management. Each strategy offers distinct advantages and is suitable for specific scenarios within a Hadoop cluster. Shared cache management involves the establishment of a centralized cache that is made accessible to multiple nodes or processes in the Hadoop cluster. Shared cache management proves particularly advantageous in scenarios where multiple jobs or tasks can benefit from a common cache, promoting efficient resource utilization and improved overall performance. On the other hand, distributed cache management focuses on maintaining separate caches on individual nodes within the Hadoop cluster. Each node possesses its dedicated cache, which is managed independently. This strategy is particularly suitable when data access patterns are localized, and each node necessitates a dedicated cache to store frequently accessed data. In contrast, centralized cache management revolves around the implementation of a single cache that is centrally managed within the Hadoop cluster. Typically, the cache resides on a dedicated server or a set of servers, facilitating access by all nodes in the cluster. This approach streamlines cache management operations and ensures consistent caching performance across the entire cluster. Centralized cache management offers the benefit of simplified administration and can contribute to improved performance by providing a unified caching mechanism for all nodes within the cluster.

Table 1 summarizes the key attributes associated with each caching mechanism as applied in Hadoop environments.

## Table 1: Caching mechanism characteristics

| Caching strategy | Hadoop cache level | Cached item | Cached item placement | Items access pattern | Cache replacement strategy | Cache management |
|---|---|---|---|---|---|---|
| **HDCache [4]** | Distributed cache | Files | Server-side | Write-once-read-many | LRU | Shared memory management |
| **HyCache [5]** | Distributed cache | Files | Server-side | Read/write | LRU/ LFU | Distributed cache management |
| **HyCache+ [6]** | Distributed cache | Files | Server-side | Read/write | LRU | Distributed cache management |
| **Haloop [8]** | Combination of HDFS cache & in-memory cache | Invariant data in the first iteration & Reduce output | Inter-iteration locality (Server-side) | Write-behind caching | LRU | Centralized cache management |
| **Incoop [9]** | Memcache | Fresh results from the previous run | Server-side | Write-through caching | LRU | Memcached management |
| **Dache [10]** | In-memory cache | tuple:{Origin, Operation} Origin is the name of a file in the DFS, and Operation is a linear list of performed operations | Server-side | Read-through caching and write-behind caching | LRU | Centralized cache management |
| **Redoop [11]** | In-memory cache at the client-side | Input data block & intermediate data block | Client-side | Read-through caching | LRU | Local cache management |
| **CurtMapReduce [12]** | Distributed cache | Results of Map tasks & Reduce Tasks | Server-side | Read-through and write-through caching | LRU | Distributed cache management |
| **Improved CL, and DL for map tasks [13]** | Distributed cache | Input data blocks for Map tasks | Server-side | Read/write | LRU | Centralized cache management |
| **LARD [14]** | Buffer cache | Input data blocks & intermediate data blocks | Server-side | Read/write | LRU | Centralized cache management |
| **CATS [15]** | Buffer cache | Input data blocks & intermediate data blocks | Server-side | Read/write | LRU | Distributed cache management |
| **ACL [16]** | Distributed cache | Input data blocks & intermediate data blocks | Server-side | Read/write | LRU | Centralized cache management |
| **CLQLMRS [17]** | Distributed cache | Input data blocks & intermediate data blocks | Server-side | Read/write | LRU | Centralized cache management |
| **Separation [18]** | HDFS cache | File system metadata | Server-side | Read/write | LRU | Centralized cache management |
| **CodedMapReduce [19]** | Distributed cache | Intermediate data blocks | Server-side | Read-modify-write approach | LRU | Distributed cache management |
| **CoARC [20]** | Distributed cache | Recently recovered data blocks | Server-side | Read/write | LRF | Centralized cache management |
| **PacMan [21]** | In- memory cache | Input Files | Server-side | Read/write | LIFE, LFU-F | Distributed cache management |
| **EDache [22]** | In-memory cache | tuple:{Origin, Operation} Origin is the name of a file in the HDFS, and Operation is a linear list of performed operations | Server-side | Read-through caching and write-behind caching | WSClock | Centralized cache management |
| **Collaborative [23]** | Distributed cache | Input data blocks & intermediate data blocks | Server-side | Read/write | Modified ARC | Centralized cache management |
| **Cache affinity aware [24]** | In-memory cache | Input data blocks | Server-side | Read/write | Cache affinity aware | Centralized cache management |
| **Block goodness aware [25]** | In-memory cache | Input data blocks | Server-side | Read/write | Block goodness aware | Centralized cache management |
| **AutoCache [27]** | In-memory cache | Input files | Server-side | Read/write | LRFU, XGB | Automated cache management |
| **H-SVM-LRU [28]** | Distributed cache | Input data blocks & intermediate data blocks | Server-side | Read/write | H-SVM-LRU | Centralized cache management |
| **Jecache [29]** | Distributed cache | Input data blocks | Server-side | Write -once-read-many | LRU | Centralized cache management |
| **Prefetch thread [30]** | Distributed cache | Input data blocks | Server-side | Read/write | LRU | Centralized cache management |
| **Streaming technique [31]** | Distributed cache | Input data blocks | Server-side | Read/write | LRU | Centralized cache management |
| **Speculative prefetch [32]** | Distributed cache | Intermediate data | Server-side | Read/write | LRU | Centralized cache management |
| **Two-level correlation-based file prefetching [33]** | Distributed cache | Input data blocks | Client-side/ server-side | Aggregate read | LRU | Centralized cache management |
| **Smart prefetch [34]** | Distributed cache | Input data blocks & intermediate data blocks | Server-side | Read/write | LRU | Centralized cache management |

## 4.2 Comparison of caching strategies

Table 2 provides a comparison of caching strategies by taking into account their techniques, advantages, disadvantages, and limitations. Moreover, it suggests appropriate scenarios to utilize these strategies.

## Table 2: Comparison of caching strategies

| Caching strategy | Technique | Advantages | Disadvantages | Limitation | Usecase |
|---|---|---|---|---|---|
| **HDCache [4]** | Distributed hash table | Improves real-time file access performance in a cloud computing environment | Does not consider complex and dynamic data access models of real-time services | Used for intranet environment outside firewall | Cloud computing environment |
| **HyCache [5]** | Middleware between HDFS and local storage, SSD | High throughput, low latency, strong consistency, and multithread support | Not scalable and only considers local data movement. | Does not support network storage and scalability via replication | Parallel file system |
| **HyCache+ [6]** | Two-layer scheduling, distributed hash table (HDT) | Minimizes network cost and scalability | Energy consumption | Does not support real-time applications | Parallel file system |
| **Haloop [8]** | Loop-aware task scheduler, loop-invariant data caching | Reduces run time and shuffling time | Requires a fixed number of Reduce tasks across iterations | A single pipeline is given in the loop instead of DAG | Iterative applications |
| **Incoop [9]** | Incremental HDFS, memorization-aware scheduler | Improves performance for applications that require frequent updates to data, reduces the amount of disk I/O and network traffic | Computational overhead in incremental HDFS small datasets | Requires additional setup and configuration | CPU-Oriented jobs, incremental applications |
| **Dache [10]** | Data-aware cache, request & reply protocol | Eliminates duplicate tasks | Cache lifetime management | Uses the same data split in both the data and cached item | Incremental processing |
| **Redoop [11]** | Adaptive window-aware data partitioning, cache-aware task scheduling | Reduces I/O cost and provides workload balancing | Increased load due to the gathering of statistical information. | Unbounded caches not controlled | Recurring query processing |
| **CurtMapReduce [12]** | Intelligent square divisor | Reduces resource wastage and computation overhead | Needs to repartition data | Not efficient for applications with little computation | Iterative applications |
| **Improved CL, and DL for map tasks [13]** | Weighted bipartite graph, maximum matching algorithm | Improves data locality rate for Map tasks | Needs scheduling time and does not consider the load of the cluster | Only considers Map tasks | Small scale cluster |
| **LARD [14]** | Predict cached data location | Suitable for small files | Poor performance for large datasets | Slow and depends on predictions | Small files |
| **CATS [15]** | Buffer cache probe to find tasks with the greatest amount of cached data | Increases cache locality rate | Only considers cache locality | Uses disk scheduling for non-cache local tasks | Realistic query execution environment |
| **ACL [16]** | Delay tasks to launch on local nodes | Increases task locality rate | Latency to schedule tasks | Scheduling overhead | High cache -affinity applications |
| **CLQLMRS [17]** | Q-Learning to train scheduling policy | Increases task locality rate | Needs to train scheduling policy | Training time | I/O- bound jobs and data-oriented applications |
| **Separation [18]** | Secondary NameNode & separation algorithm | Name node scalability | Needs to determine a threshold value | Stores only the most frequent data | Decreasing NameNode load & frequent data access |
| **CodedMapReduce [19]** | Create coding opportunities | Reduces inter-server communication for shuffling | Additional processing time to compute intermediate results | A trade-off between computation load & communication load | Repetitive execution of Map tasks |
| **CoARC [20]** | Data recovery mechanism | Reduces network traffic | Only reduce Map time | Only considers Map tasks | Data recovery in Map tasks |
| **PacMan [21]** | Frequency access and uncompleted file | Minimizes job completion time & maximizes cluster efficiency | Uses all-or-nothing property | Small jobs not supported | Data-intensive parallel jobs |
| **EDache [22]** | Last time used | Decreases execution time | Long search times for large data blocks | Uses the same data split for both the data and cached items | Incremental processing |
| **Collaborative [23]** | Recency and frequency of access | Increases cache hit ratio | Needs space for storing history | Memory limitations | Data-local jobs |
| **Cache affinity aware [24]** | Cache affinity of application and recency | Effective use of cache space | Needs to calculate cache affinity of applications | Needs to know application cache affinity | Applications with high cache affinity |
| **Block goodness aware [25]** | Block goodness value and access time | Effective for multiple concurrent applications | Needs to calculate block goodness | Needs to know block goodness | Static workloads |
| **AutoCache [27]** | Light-weight gradient boosted trees for learning file access patterns | Reduces average completion time and improves cluster efficiency | File oriented cache | Depends on machine learning model | Data processing workloads |
| **H-SVM-LRU [28]** | Combination of SVM and LRU | Increases cache hit ratio | Training time | Need for training data | I/O-bound jobs and high cache affinity applications |
| **Jecache [29]** | Just in time block prefetching | Uses cache space efficiently | Needs to measure the average processing time of data blocks | Only considers read caching | I/O-bound workloads |
| **Prefetch thread [30]** | Prefetch data | Reduces transmission time | Waiting time for first data transmission | Only considers Map tasks | I/O-bound workloads |
| **Streaming technique [31]** | Streaming technique | Reduces waiting time | No load balance considerations | Only considers Map tasks | I/O-bound workloads |
| **Speculative prefetch [32]** | Speculative prefetch & KNN clustering | Improves data locality rate and execution time | Does not take into account node throughput | Needs to determine the number of clusters (K) | I/O-bound workloads |
| **Two-level correlation-based file prefetching [33]** | Dynamic replica selection | Reduces access latency | Does not consider several features | Depends on access patterns | Hadoop-based Internet applications |
| **Smart prefetch [34]** | Calculates suitable prefetch time, and prefetched data volume using KNN | Increases local task rate | Needs to calculate task progress rate | Depends on threshold value | High cache affinity applications, I/O-bound jobs |

## 4.3 The impact of caching strategies on Hadoop performance

In this section, we investigate the effects of the presented caching strategies on Hadoop performance by considering various performance metrics: data access time, job execution time, resource utilization, scalability, overhead, and data locality.

## Table 3: Impact of caching strategies on Hadoop performance

| Caching strategy | Data access time | Job execution time | Resource utilization | Scalability | Overhead | Data locality |
|---|---|---|---|---|---|---|
| **HDCache [4]** | Reduces disk accesses | Reduces execution time by improving data access time | N/A | No | Omits network traffic overheads | N/A |
| **HyCache [5]** | Accelerates HDFS | Improves execution time | N/A | No | High | N/A |
| **HyCache+ [6]** | Accelerates HDFS | Speed up by 29X in their tests | N/A | Yes | High | Achieves data locality using the novel 2LS |
| **Haloop [8]** | Improves invariant data access without recomputation | Reduces run time | Avoids recomputation of invariant data between loops | No | Overhead of shuffling invariant data is completely avoided | Inter-iteration locality |
| **Incoop [9]** | Reduces intermediate data access time by reusing results from previous runs | Reduces execution time via location-aware scheduling and minimizes data transmission across networks | Avoids recomputation, prevents idle nodes | N/A | Computational overhead in incremental HDFS for small datasets | Location-aware scheduling for memorization |
| **Dache [10]** | Improves access to intermediate data by caching them and their operations | Improves completion time | Improves CPU utilization | No | Cache request and reply protocol avoids network communication | N/A |
| **Redoop [11]** | Improves access to overlapped data by caching intermediate data | Improves execution time by avoiding redundant computations | Reduces I/O costs by avoiding unnecessary reloading, reshuffling, and re-computation of overlapping data | Yes | Gathering of statistical information | Cache locality |
| **CurtMapReduce [12]** | Access intermediate results | Improves execution time by avoiding wastage of resources | Prevents resource wastage by eliminating duplicate processing | Yes | Avoids significant overhead | N/A |
| **Improved CL, and DL for map tasks [13]** | Improves input data access time by caching them | Improves execution time by increasing data localization usage | Selects appropriate resources based on weighted bipartite graph maximum match | No | N/A | Data, cache locality (Map tasks) |
| **LARD [14]** | Improves input data access time by caching them | Improves execution time by increasing cache locality rate | N/A | No | N/A | Cache locality |
| **CATS [15]** | Improves input data access time by caching them | Improves execution time by increasing cache locality rate | N/A | Yes | Reduces overhead | Cache locality |
| **ACL [16]** | Improves input data access time by caching them | Improves execution time by increasing local task rate | N/A | No | Waiting time to schedule by skipping some nodes to launch local task | Data, cache locality |
| **CLQLMRS [17]** | Improves input data access time by caching them | Improves execution time by increasing local task rate | N/A | Yes | Decreases overhead by using reinforcement learning | Data, cache locality |
| **Separation [18]** | Accesses more metadata | Improves job execution time due to increased metadata availability | N/A | NameNode Scalability | Increases overhead by storing more metadata | N/A |
| **CodedMapReduce [19]** | Improves by coded caching | Improves execution time by reducing shuffling time | N/A | N/A | Coded calculation | N/A |
| **CoARC [20]** | Improves access to unavailable data blocks | Reduces execution time by eliminating redundant recovery and reducing Map time | Reduces network usage by recovering temporarily unavailable blocks in the same strip | Yes | Recovers unavailable data blocks without any increase in network traffic | N/A |
| **PacMan [21]** | Improves frequently accessed files | Reduces job completion times via LIFE | Maximizes efficiency of the cluster via LFU-F | Yes | N/A | Memory locality |
| **EDache [22]** | Improves access to intermediate data by caching them and their operations | Improves completion time | Improves CPU utilization | No | Cache request and reply protocol avoids network communication | N/A |
| **Collaborative [23]** | Improves data access time via modified ARC | Decreases execution time by increasing cache hit ratio | N/A | Yes | N/A | Data locality |
| **Cache affinity aware [24]** | Improves data access time by caching them | Improves execution time by considering cache affinity features | N/A | No | Reduces scheduling overhead | Data locality, cache locality |
| **Block goodness aware [25]** | Improves data access time via block goodness aware cache replacement | Improves execution time by increasing cache hit ratio | Improves cache utilization | Yes | Uses block granularity caching with a small overhead | Cache locality |
| **AutoCache [27]** | Improves data access time by caching them | XGB reduces completion time | Cluster efficiency | No | Negligible CPU overhead caused by XGBoost for both training and prediction | Temporal locality |
| **H-SVM-LRU [28]** | Improves data access time by avoiding cache pollution | Improves job execution time by increasing cache hit ratio | Improves cache utilization by avoiding cache pollution | No | Needs training data | N/A |
| **Jecache [29]** | Improves data access time by just-in time data prefetching | Reduces job execution time by accessing cached input data before processing them | Improves memory cache utilization | Yes | Less extra overhead since less threads need to be initiated to prefetch blocks | N/A |
| **Prefetch thread [30]** | Improves data access time by prefetching them | Decreases execution time by fetching data in advance | Improves I/O utilization | No | N/A | N/A |
| **Streaming technique [31]** | Improves data access time by prefetching them | Improves execution of Map tasks | Improves I/O utilization | Yes | Packet overhead problem | N/A |
| **Speculative prefetch [32]** | Improves data access time by prefetching them | Reduces execution time via KNN clustering | Improves I/O utilization | Yes | Hides the overhead involved during data transmission | Data locality |
| **Two-level correlation-based file prefetching [33]** | Reduces access latency on the metadata server | Improves execution time by eliminating access delay | Improves CPU utilization of NameNode | No | Reduces network transfer overhead | Data locality |
| **Smart prefetch [34]** | Improves data access time by prefetching at a suitable time | Improves execution time by increasing data locality rate | Improves I/O utilization | Yes | Extra overhead to calculate tasks' progress | Data locality |

## 4.3 Discussion and research opportunities

This section presents a discussion and comparison of various caching approaches employed in the context of Hadoop. Statistical information, as illustrated in Figure 1, is provided to compare the papers reviewed in terms of their contributions to addressing Hadoop-related issues. As previously mentioned in Section 3, the caching methods are categorized into two groups: basic and enhanced. It is evident that researchers have conducted more studies on enhanced caching techniques in order to harness the additional benefits offered by caching mechanisms. Notably, the topic garnering the highest popularity among these studies is the improvement of cache replacement algorithms, which aims to increase the cache hit ratio. This particular area of research accounts for approximately 26% of all studies. The avoidance of recomputation and the reduction of data transmission are next in popularity, each constituting nearly 16% of all studies.

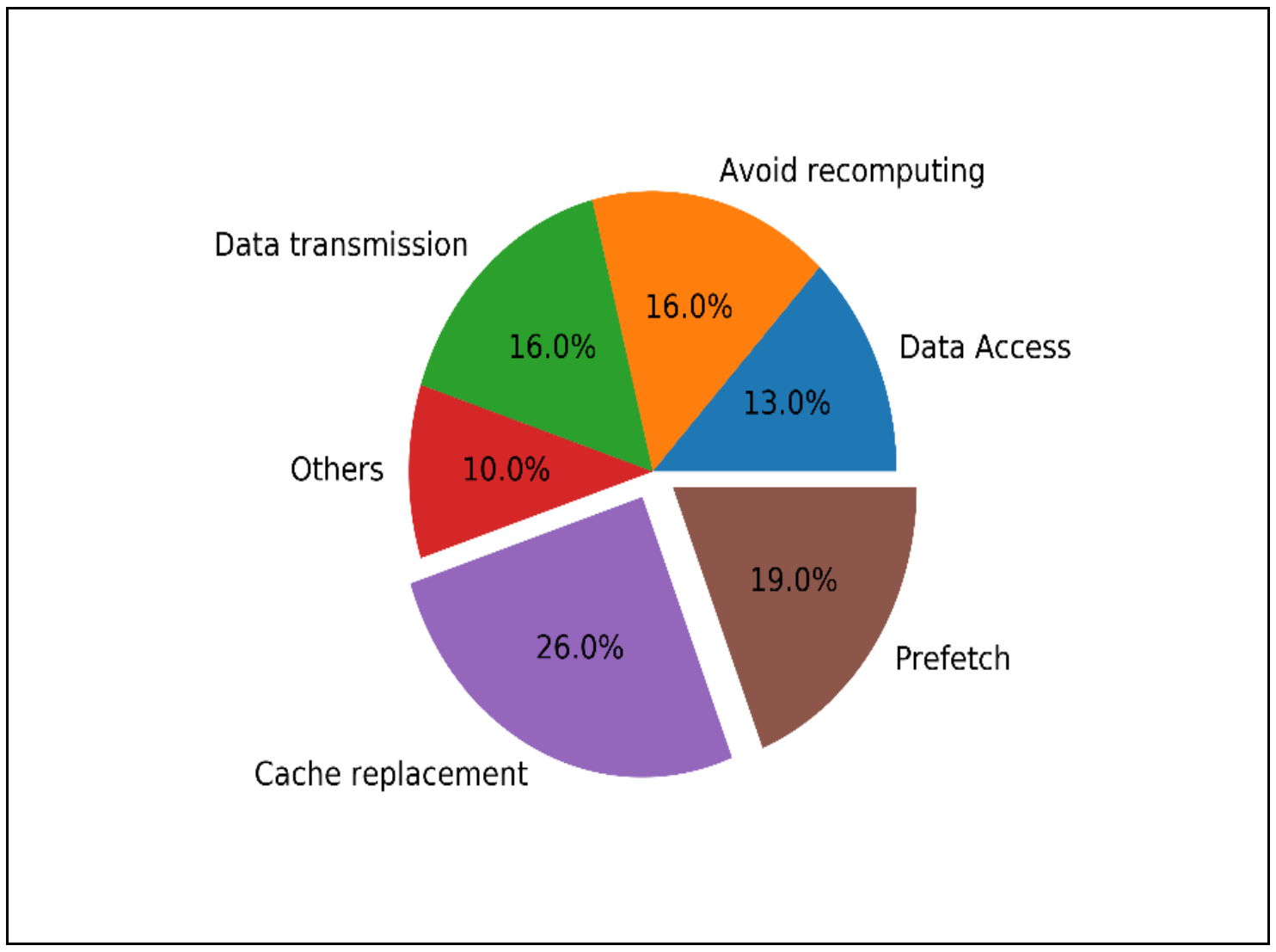

**Figure 1: Percentage of the papers reviewed in terms of the problem-solving approach**

Furthermore, Figure 2 presents the distribution of performance parameters used for evaluating caching mechanisms. The analysis reveals that execution time is the most significant performance metric, accounting for 47% of the evaluations. Data locality is next with a share of 15%. The prominence of these two metrics can be attributed to their substantial impact on overall performance. Conversely, overhead has the lowest share among these parameters, underscoring the need for greater attention to this aspect in future research endeavors.

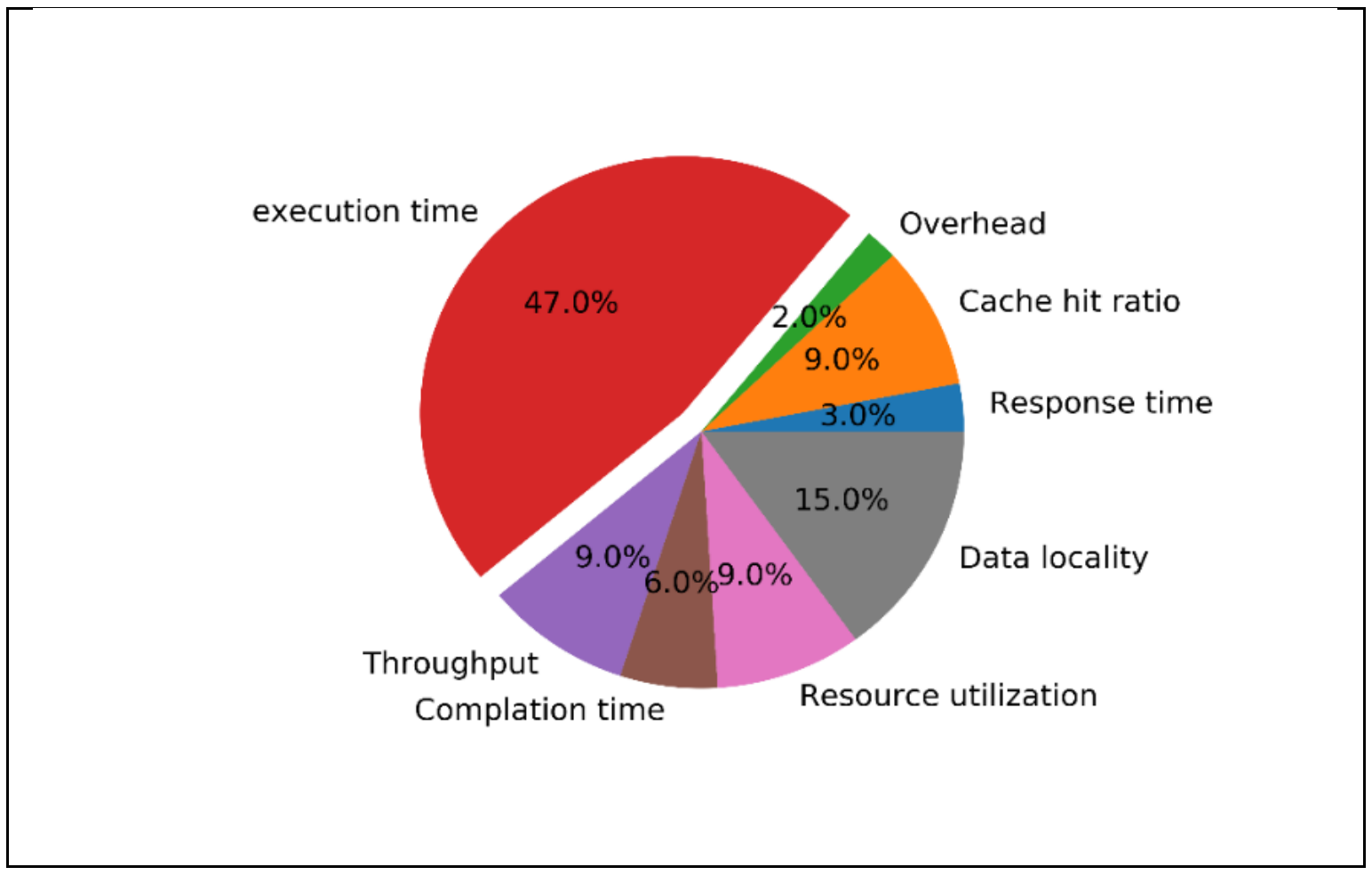

**Figure 2: Percentage of performance metrics for evaluating caching algorithms**

Based on the findings of this survey, it is evident that the caching mechanism in Hadoop remains a significant area of research interest, necessitating further investigation. As a prospective avenue for exploration, novel caching strategies could be devised to optimize multiple metrics simultaneously (or manage trade-offs between them), thereby addressing and resolving various challenges encountered in Hadoop. For instance, caching mechanisms can be employed to tackle multiple difficulties in Hadoop simultaneously, or a hybrid approach combining two methods can be developed to enhance performance. These avenues present promising opportunities for future research. In the subsequent section, we propose a hybrid method that combines CLQLMRS and H-SVM-LRU, taking into consideration job execution time as a performance metric for evaluation.

## 5. Hybrid approach

We investigate basic and enhanced caching mechanisms and propose a novel hybrid approach called Hybrid Intelligent Cache (HIC) that combines the benefits of both methods. The HIC framework incorporates the CLQLMRS MapReduce job scheduler, which employs reinforcement learning to train scheduling policies considering data and cache locality for task assignment to worker nodes. The CLQLMRS scheduler improves task locality rates, resulting in reduced data transmission time and enhanced Hadoop performance. Additionally, the HIC framework incorporates the H-SVM-LRU intelligent cache replacement strategy, which optimizes cache space utilization by preventing cache pollution. This strategy classifies cached items into reusable and non-reusable groups and determines eviction based on item class and the Least Recently Used (LRU) policy. The proposed hybrid approach is evaluated through experiments conducted on a Hadoop cluster to assess its impact on system performance metrics.

### 5.1 Experimental setup

For carrying out the experiment, we use a cluster consisting of a single NameNode and nine DataNodes located in two racks such that odd-numbered nodes are in rack1 and even-numbered nodes are in rack2.

- *Hardware configuration*: The nodes are connected via a 10 Gigabit Ethernet switch. The experimental environment is heterogeneous with different memory sizes at the nodes: NameNode has16 GB RAM; DataNode 1, DataNode 4, and DataNode 7 have 4 GB RAM, DataNode 2, DataNode 5, and DataNode 8 have 6 GB RAM, DataNode 3, DataNode 6, and DataNode 9 have 8 GB RAM. The CPU for all nodes is an Intel Core i7-6700 processor and a one TB hard disk.
- *Software configuration*: We use the Ubuntu14.04 operating system and JDK 1.8, Hadoop version 2.7 (which employs in-memory caching), and Intel HiBench [35] [36] version 7.1 as the Hadoop benchmark.
- *Hadoop configuration parameters:* The block size of files in HDFS is 128 MB, the number of cache replicas is set to one, and data replication is set to 3. The memory sizes for Map tasks, Reduce tasks, and the node manager are 1GB, 2GB, and 8 GB, respectively. The maximum size of the cache is set to 1.5 GB and we assume that each DataNode in the cluster has the same size cache. The remaining Hadoop configuration parameters are set to default values.
- *MapReduce applications*: We use Intel HiBench as a Hadoop benchmark suite and carry out our experiments by using two groups of benchmarks: 1) Micro benchmarks that contain

WordCount, Sort, and TeraSort applications. 2) Hive benchmarks as a big data query and analysis application, for instance, Join and Aggregation. WordCount is a CPU-intensive application that counts occurrences of each word in a text file. Sort is a typical I/O-bound application with moderate CPU usage and sorts input data. TeraSort is an I/O-oriented program that needs moderate disk I/O during the Map and Shuffle phases and heavy disk I/O in the Reduce phase. Both Aggregation and Join are supported by Hive and used for operation in the query. Join is a multiple-stage application where the results of each step are used as input for the next step.

- *Input data:* For carrying out experiments, we have used the default data sizes from the HiBench suite. Sort and WordCount have 60 GB and TeraSort has 1 TB as input data size. The input data for WordCount, Sort, and TeraSort are generated by using the RandomTextWriter, RandomWriter, and TeraGen programs respectively, all contained in the Hadoop distribution.
- *Metric:* We consider job execution time as the metric to evaluate our purposed method. It plays a vital role in Hadoop performance improvement and is related to data access time. The data access time decreases significantly if we could access data from the cache instead of the disk, reducing the job execution time. To calculate the average job execution time, we run each application five times.
- *Scenarios:* In this experiment, we take into account four scenarios. First, Hadoop original does not utilize HDFS in-memory caching and used the default MapReduce job scheduler (FIFO). We use this as a baseline. Second, we utilize H-SVM-LRU instead of LRU as a cache replacement algorithm. Next, we apply CLQLMRS as a job scheduler to train a scheduling policy that considers both data locality and cache locality. Finally, we compose H-SVM-LRU with CLQLMRS as a hybrid method.

### 5.2 Evaluation

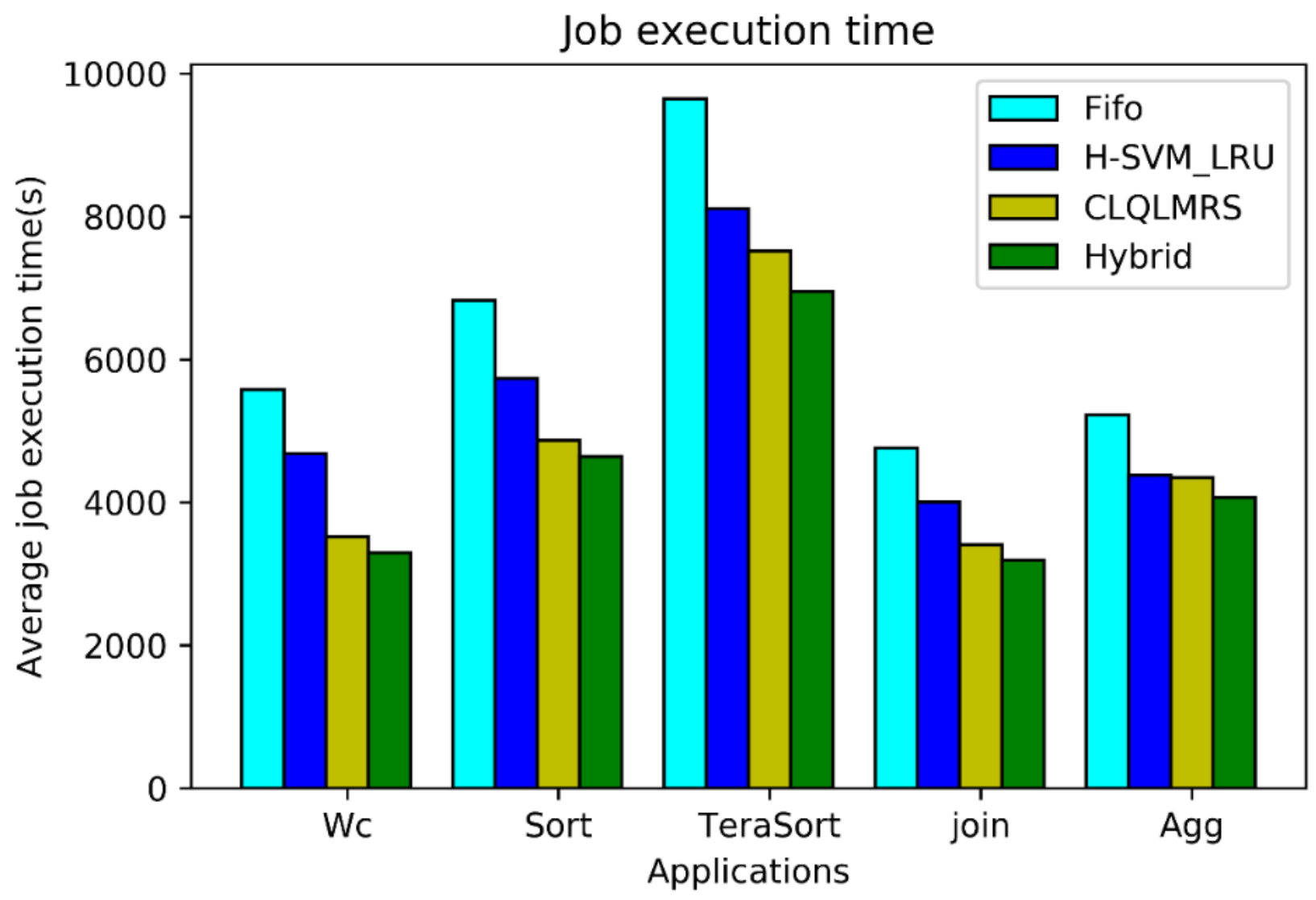

**Figure 3: Average job execution time for various applications**

The average job execution time for the considered applications is depicted in Figure 3. The results demonstrate that the H-SVM-LRU approach exhibits a notable improvement in job execution time, achieving a 16% reduction compared to the Hadoop-original configuration. This improvement can be attributed to H-SVM-LRU's ability to effectively utilize the limited cache space and mitigate cache pollution, resulting in an enhanced cache-hit ratio. Consequently, a larger number of tasks can leverage cached input data, leading to reduced data access time and improved overall performance. Furthermore, the CLQLMRS technique, which considers both cache locality and data locality in its scheduling policy, displays significant enhancements in the job execution time, with an approximate improvement of 26.54% compared to the FIFO approach. By prioritizing tasks that benefit from cache and data locality, CLQLMRS optimizes task allocation, resulting in more efficient execution and reduced latency. Building upon the advantages of both CLQLMRS and H-SVM-LRU, our proposed hybrid method combines these approaches to further enhance job execution time. By leveraging the benefits offered by both techniques, our hybrid method achieves an average improvement of 31.2% in job execution time across the evaluated applications. This demonstrates the potential of integrating cache locality, data locality, and advanced cache replacement strategies in optimizing job execution and resource allocation for improved performance.

## 6. Conclusion

This paper examines the utilization of caching mechanisms as a potential solution to address various challenges within the Hadoop framework. The study encompasses an analysis of two primary categories of caching mechanisms: basic caching and enhanced caching. Subsequently, a novel classification scheme is introduced to categorize basic caching strategies based on the specific problem they seek to alleviate, such as enhancing data access time, eliminating redundant computations, reducing data transmission, and other related objectives. Enhanced caching is further subdivided into two groups: improved cache replacement algorithms and enhanced prefetch mechanisms. Each method is described, and a comparative analysis is conducted to evaluate their respective advantages, disadvantages, and impact on the overall performance of Hadoop. Moreover, the paper presents a novel hybrid intelligent caching approach that leverages the combined benefits of the H-SVM-LRU cache replacement algorithm and the CLQLMRS scheduling policy. Experimental findings demonstrate that this novel approach yields a notable improvement of 31.2% in execution time. This enhancement is attributed to the heightened cache hit ratio facilitated by the H-SVM-LRU cache replacement algorithm, as well as the increased likelihood of tasks utilizing local data due to the CLQLMRS scheduling policy. In light of these achievements, we are interested in exploring additional machine learning techniques to further investigate the potential of intelligent caching. We have a goal of expanding the application of machine learning methodologies to advance the field of caching mechanisms within Hadoop, by addressing more intricate challenges.

## References

1. “Apache Hadoop” http://Hadoop. apache. org/

2. J. Dean and S. Ghemawat. MapReduce: Simplified Data Processing on Large Clusters. In OSDI, pages 137–150, (2008)

3. K. Shvachko, H. Kuang, S. Radia, and R. Chansler, The Hadoop Distributed File System, IEEE 26th Symposium on Mass Storage Systems and Technologies (MSST), (2010)

4. Zhang, J., Wu, G., Hu, X. & Wu, X. A distributed cache for Hadoop Distributed File System in real-time cloud services. Proceedings - IEEE/ACM International Workshop on Grid Computing 12–21 (2012) doi:10.1109/Grid.2012.17.

5. Zhao, D. & Raicu, I. HyCache: A user-level caching middleware for distributed file systems. Proceedings - IEEE 27th International Parallel and Distributed Processing Symposium Workshops and Ph.D. Forum, IPDPSW 2013 1997–2006 (2013) doi:10.1109/IPDPSW.2013.83.
6. Zhao, D., Qiao, K. & Raicu, I. HyCache+: Towards scalable high-performance caching middleware for parallel file systems. Proceedings - 14th IEEE/ACM International Symposium on Cluster, Cloud, and Grid Computing, CCGrid 2014 267–276 (2014) doi:10.1109/CCGrid.2014.11.
7. Singh, G., Chandra, P. & Tahir, R. A Dynamic Caching Mechanism for Hadoop using Memcached, (2012).
8. Bu, Y., Howe, B. & Ernst, M. D. HaLoop : Efficient Iterative Data Processing on Large Clusters. 3, (2010).
9. Bhatotia, P., Wieder, A., Rodrigues, R., Acar, U. A. & Pasquini, R. Incoop: MapReduce for incremental computations. Proceedings of the 2nd ACM Symposium on Cloud Computing, SOCC 2011 (2011) doi:10.1145/2038916.2038923.
10.Yaxiong Z., Jie W., & Cong L. Dache: A Data Aware Caching for Big-Data Applications Using the MapReduce Framework" Tsinghua Science and Technology ISSNl l1007-0214l l05/10l l pp39-50 Volume 19, Number 1, February (2014).
11. Lei, C., Zhuang, Z., Rundensteiner, E. A. & Eltabakh, M. Y. Redoop infrastructure for recurring big data queries. Proceedings of the VLDB Endowment 7, 1589–1592 (2014).
12. Huang, T. C., Chu, K. C., Zeng, X. Y., Chen, J. R. & Shieh, C. K. CURT MapReduce: Caching and utilizing results of tasks for MapReduce on cloud computing. Proceedings - 2016 IEEE 2nd International Conference on Multimedia Big Data, BigMM 2016 149–154 (2016) doi:10.1109/BigMM.2016.10.
13. Zhang, P., Li, C. & Zhao, Y. An improved task scheduling algorithm based on cache locality and data locality in Hadoop. Parallel and Distributed Computing, Applications and Technologies, PDCAT Proceedings 0, 244–249 (2016).
14. Pai, V.S., Aron, M., Banga, G., Svendsen, M., Druschel, P.Zwaenepoel, W., Nahum, E.: Locality-aware request distribution in cluster-based network servers. In: ACM Sigplan Notices, vol. 33, pp. 205–216. ACM (1998)
15. Lim, B., Kim, J. W. & Chung, Y. D. CATS: cache-aware task scheduling for Hadoop-based systems. Cluster Computing 20, 3691–3705 (2017).
16. Floratou, A. et al. Adaptive caching in Big SQL using the HDFS cache. Proceedings of the 7th ACM Symposium on Cloud Computing, SoCC 2016 321–333 (2016) doi:10.1145/2987550.2987553.
17. Ghazali, R., Adabi, S., Rezaee, A., Down, D. G. & Movaghar, A. CLQLMRS: improving cache locality in MapReduce job scheduling using Q-learning. *Journal of Cloud Computing* 11, (2022).
18. Mukhopadhyay, D., Agrawal, C., Maru, D., Yedale, P. & Gadekar, P. Addressing name node scalability issue in Hadoop distributed file system using cache approach. Proceedings - 2014 13th International Conference on Information Technology, ICIT 2014 321–326 (2014) doi:10.1109/ICIT.2014.18.
19. Li, S., Maddah-Ali, M. A. & Avestimehr, A. S. Coded MapReduce. 1–16 (2015).
20. Subedi, P. et al. CoARC: Co-operative, Aggressive Recovery and Caching for Failures in Erasure Coded Hadoop. Proceedings of the International Conference on Parallel Processing 2016-September, 288–293 (2016).
21. Ananthanarayanan, G., Ghodsi, A., Wang, A.,Borthakur, D., Kandula, S., Shenker, S., Stoica, I., Pacman: Coordinated Memory Caching for Parallel Jobs" University of California, Berkeley, Facebook, Microsoft Research, KTH/Sweden.
22. Sangavi, S. *et al.* An Enhanced DACHE model for the MapReduce Environment. *Procedia - Procedia Computer Science* 50, 579–584 (2015).
23. Shrivastava, M. & Bischof, H. Hadoop-Collaborative Caching in Real-Time HDFS, thesis, Rochester Institute of Technology Rochester, New York. (2012).
24. Code, L. Master's Thesis Cache Affinity-aware In-memory Caching Management for Hadoop Jaewon Kwak.
25. 13. Kwak, J., Hwang, E., Yoo, T., Nam, B. & Choi, Y. In-memory Caching Orchestration for Hadoop. (2016) doi:10.1109/CCGrid.2016.73.
26. A. Floratou, N. Megiddo, N. Potti, F. Ozcan, U. Kale, and J. Schmitz-Hermes. Adaptive Caching in Big SQL using the HDFS Cache. In Proc. of the 7th ACM Symp. on Cloud Computing (SoCC). ACM, 2016.
27. Herodotou, H. AutoCache: Employing machine learning to automate caching in distributed file systems. *Proceedings - 2019 IEEE 35th International Conference on Data Engineering Workshops, ICDEW 2019* 133–139 (2019) doi:10.1109/ICDEW.2019.00-21.
28. Ghazali, R., Adabi, S., Rezaee, A., Down, D. G. & Movaghar, A. Hadoop-Oriented SVM-LRU (H-SVM-LRU): An Intelligent Cache Replacement Algorithm to Improve MapReduce Performance. (2023).
29. Luo, Y., Shi, J. & Zhou, S. JeCache: Just-Enough Data Caching with Just-in-Time Prefetching for Big Data Applications. Proceedings - International Conference on Distributed Computing Systems 2405–2410 (2017) doi:10.1109/ICDCS.2017.268.

30. Vinutha, D. C. & Raju, G. T. Data Prefetching for Heterogeneous Hadoop Cluster. 2019 5th International Conference on Advanced Computing and Communication Systems, ICACCS 2019 554–558 (2019) doi:10.1109/ICACCS.2019.8728373.
31. Lee, J., Kim, K. T. & Youn-Chen, T. MapReduce Performance Scaling Using Data Prefetching, 9, 26–31 (2022).
32. Kalia, K. et al. Improving MapReduce heterogeneous performance using KNN fair share scheduling. Robotics and Autonomous Systems 157, 104228 (2022).
33. Dong, B. et al. Correlation-based file prefetching approach for Hadoop. Proceedings - 2nd IEEE International Conference on Cloud Computing Technology and Science, CloudCom 2010 41–48 (2010) doi:10.1109/CloudCom.2010.60.
34. Ghazali, R. & Down, D. G. Smart Data Prefetching Using KNN to Improve Hadoop Performance. 1–18 (2023).
35. Huang S, Huang J, Dai J, Xie T, Huang B The HiBench Benchmark Suite: Characterization of the MapReduce-Based Data Analysis. https://doi. org/ 10. 1109/ ICDEW. 2010. 54527 47 (2014).
36. "Hibench" https://github.com/ Intel- bigdata/ HiBench